\documentclass[twocolumn]{aastex631}

\usepackage{amsfonts}
\usepackage{amsmath}
\usepackage{amssymb}
\usepackage{subfigure}
\usepackage{graphicx}
\usepackage{epstopdf}
\usepackage{epsfig}
\usepackage{float}
\usepackage{hyperref}
\usepackage{color}
\usepackage{ulem}
\usepackage{cancel}
\usepackage{mathrsfs}
\usepackage{booktabs}
\usepackage{multirow}
\usepackage{natbib}
\usepackage{ulem}

\begin{document}

\title{Fast Radio Bursts with Narrow Beaming Angles Can Escape from Magnetar Magnetospheres}

\author[0009-0009-7749-8998]{Yu-Chen Huang}
\affiliation{Department of Astronomy, University of Science and Technology of China, Hefei 230026, China; daizg@ustc.edu.cn}
\affiliation{School of Astronomy and Space Science, University of Science and Technology of China, Hefei 230026, China}

\author[0000-0002-7835-8585]{Zi-Gao Dai}
\affiliation{Department of Astronomy, University of Science and Technology of China, Hefei 230026, China; daizg@ustc.edu.cn}
\affiliation{School of Astronomy and Space Science, University of Science and Technology of China, Hefei 230026, China}

\begin{abstract}

	Fast radio bursts (FRBs) are millisecond duration transients observed in the radio band, with their origin and radiation mechanism remaining unclear to date. Growing evidence indicates that at least some FRBs originate from magnetars and are likely generated within the magnetospheres of these highly magnetized neutron stars. However, a recent study suggested that FRBs originating from magnetar magnetospheres would be scattered by magnetospheric electron--positron pair plasma, making it impossible for them to escape successfully. In this paper, we first demonstrate that the scattering effect can be greatly attenuated if the angle between the FRB propagation direction and the background magnetic field is $\sim10^{-2} \text{ rad}$ or smaller. When the angle is around $10^{-1} \text{ rad}$, the beaming effect of FRBs becomes significant in reducing scattering. Such FRBs have small transverse spatial sizes, which can help them instantly push the front plasma laterally out of the radiation region. This significantly mitigates the FRB-induced two-photon annihilation reaction, $\gamma+\gamma\to e^{-}+e^{+}$, which was previously regarded as a key factor hindering the propagation of FRBs. A critical radiation cone half-opening angle between $10^{-3}-10^{-2}\text{ rad}$ is found for an FRB with isotropic luminosity $L_{\text{iso}}\sim 10^{42}\text{ erg s}^{-1}$ and emitted at a radius $r_{\text{em}}\lesssim 10^9\text{ cm}$ in the magnetosphere of a magnetar. Smaller beaming angles and larger emission radii can be more advantageous for the propagation of FRBs in magnetospheres. Our result supports the scenario that FRBs could originate from magnetar magnetospheres.

\end{abstract}



\section{Introduction}

It has been seventeen years since the first fast radio burst (FRB) was discovered \citep{Lorimer2007}. Despite significant achievements in this field, many problems remain unresolved, including the origin and radiation mechanism of FRBs. The magnetar origin of at least some FRBs has been confirmed by the association of FRB 200428 with the galactic magnetar SGR 1935+2154 \citep{Bochenek2020,CHIME/FRBCollaboration2020}. Nevertheless, the inferred isotropic peak luminosities of the bursts detected from SGR 1935+2154 are apparently smaller than those of bursts at cosmological distances \citep{Zhang2020,Kirsten2021,Dong2022,Pearlman2022}. The radiation mechanism of FRBs is also puzzling. Releasing a large amount of energy within one millisecond implies that their radiation is highly coherent, which may be somewhat challenging to understand.

Various radiation models have been proposed based on the hypothesis that FRBs originate from neutron stars. In general, these models can be broadly classified into two main categories: magnetospheric models \citep{Kumar2017,Yang2018,Lu2020,Wang2022,Zhang2022,Liu2023,Qu2024} and far-away models \citep{Lyubarsky2014,Beloborodov2017,Beloborodov2020,Metzger2019,Margalit2020}. Accumulated observations indicate that FRBs share some similarities with pulsars, which therefore favors their magnetospheric origin. Some potential observational evidence includes: frequency drifting \citep{CHIME/FRBCollaboration2019,Hessels2019,Pleunis2021,Zhou2022}, evolution of polarization position angles \citep{Cho2020,Luo2020,Mckinven2024,Niu2024}, sub-second periodicity \citep{Chime/FrbCollaboration2022}, nanosecond temporal variability \citep{Majid2021,Nimmo2022}, and scintillation \citep{Nimmo2024}, etc. Recently, some discussions on the polarization property \citep{Wang2022a,Qu2023,Jiang2024,Liu2024,Zhao2024}, narrow spectrum \citep{Yang2023,Kumar2024,Wang2024}, and waiting-time distribution \citep{Xiao2024} of FRBs also provide clues to the magnetospheric origin.

However, a potential issue for magnetospheric origin is that FRBs could be scattered by a magnetospheric dense plasma, which makes it difficult for them to escape \citep{Beloborodov2021,Beloborodov2022,Beloborodov2023}. This can be attributed to two reasons. The first reason is the enhancement of the scattering cross section due to the strong wave effect of FRBs. The second reason, perhaps more importantly, is that the plasma accelerated by an FRB pulse can emit high-energy gamma photons through curvature radiation. These photons collide with each other and generate considerable electron--positron plasma through the reaction $\gamma+\gamma\to e^{-}+e^{+}$, thereby boosting the plasma number density. Although subsequent studies pointed out that the scattering cross section of particles can be reduced if the angle between the FRB propagation direction and the local magnetic field is small or the pre-wave plasma has an extremely relativistic outflow \citep{Qu2022,Lyutikov2024}, the issue of additional electron--positron-pair creation has not been thoroughly solved yet so far. Therefore, whether FRBs can escape from magnetar magnetospheres still remains a topic which is worthy of research.

This paper is organized as follows.  In Section \ref{scmm}, we briefly summarize the scattering issue of FRBs in magnetar magnetospheres. In Section \ref{frbpp}, we reinvestigate the motion of a magnetospheric particle under the influence of an FRB by generalizing the result of \cite{Beloborodov2021,Beloborodov2022} to the case of oblique propagation. The Lorentz factor of the particle is boosted by the FRB, leading to an enhancement in the scattering cross section. On the one hand, the particle has a nearly constant velocity component along the direction of the background magnetic field, depending on the angle between the FRB propagation direction and the field line. On the other hand, the particle exhibits significant gyration around the field line. Such gyration is responsible for producing a large number of high-energy photons that contribute to the reaction $\gamma+\gamma\to e^{-}+e^{+}$. In Section \ref{approaches}, we discuss several approaches to reduce the scattering effect. We find that: (1) When the angle between the FRB propagation direction and the background magnetic field line is $\theta\lesssim10^{-2} \text{ rad}$, pair creation can be effectively inhibited. (2) For an FRB with a small transverse spatial size, the pair creation can also be significantly reduced. This effect becomes significant primarily when the propagation angle $\theta\sim10^{-1}\text{ rad}$. Based on this result, we then constrain the beaming angle and emission radius of FRBs in stellar magnetospheres. The main discussion is presented in Section \ref{discussionsec}. We adopt the notation $Q_n=Q/10^n$ and cgs units in the whole paper.

\section{Scattering of FRBs in Magnetar Magnetospheres}\label{scmm}

\subsection{Scattering Cross Section}

\begin{figure}
	\begin{center}
		\subfigure{
			\includegraphics[width=0.35\textwidth]{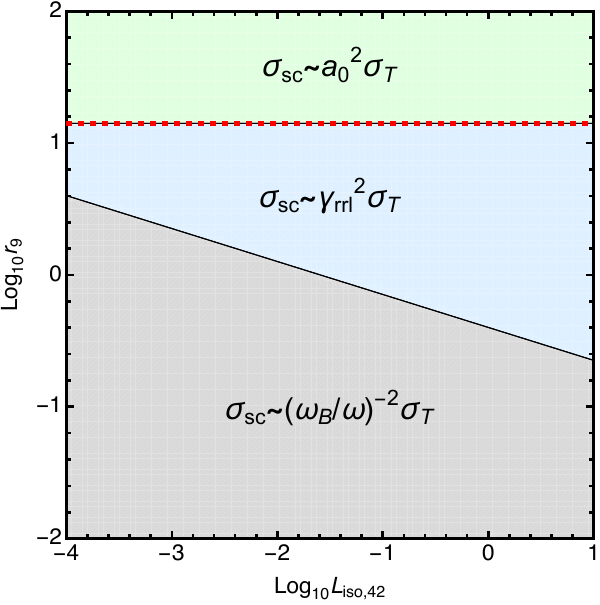}}
		\caption{Electron scattering cross sections in different regions of magnetosphere parameter space. The $x-$ and $y-$ axes represent the FRB isotropic luminosity and the distance from the magnetar, respectively. The regions where $1<a_0<\omega_B/\omega$, $1<\omega_B/\omega<a_0$, and $\omega_B/\omega<1<a_0$ are marked in gray, blue, and green, respectively. The red dashed line represents the light cylinder radius, which also serves as the boundary between the blue and green regions. Scattering is most severe in the blue region, while it is relatively weak in the gray and green regions. The parameters $\nu=10^9\text{ Hz}$, $B_{\text{s}}=10^{15}\text{ G}$, and the magnetar spin period $P=2.95\text{ s}$ are adopted.}
		\label{scacrosec}
	\end{center}
\end{figure}

The scattering of an electromagnetic wave by an electron can be typically described by the Thomson cross section \citep{Rybicki1991}
\begin{equation}
	\sigma_T=\frac{8\pi}{3}r_e^2\approx6.7\times10^{-25}\text{ cm}^2,
	\label{thomsonsc}
\end{equation}
where $r_e=e^2/mc^2$ is the classical electron radius. Here, $e$, $m$, and $c$ denote electron charge, mass, and the speed of light, respectively. For an electron inside an FRB pulse within a magnetar magnetosphere, the Thomson cross section in Equation (\ref{thomsonsc}) becomes inaccurate for two reasons. Firstly, the relativistic effect of the electron motion induced by such a large amplitude electromagnetic wave is significant. Secondly, the strong background magnetic field in the magnetar magnetosphere further complicates the situation. To quantify these effects on scattering, one can define two dimensionless parameters,
\begin{equation}
	\begin{aligned}
		a_0&\equiv\frac{eE_0}{mc\omega}=\frac{e}{mc\omega}\sqrt{\frac{2L_{\text{iso}}}{c}}\left(\frac{1}{r}\right),\\
		\frac{\omega_B}{\omega}&\equiv\frac{eB_{\text{bg}}}{mc\omega}=\frac{eB_{\text{s}}R_{\text{s}}^3}{mc\omega}\left(\frac{1}{r}\right)^3,
		\label{a0omegab}
	\end{aligned}
\end{equation}
where $E_{0}$, $L_{\text{iso}}$, and $\omega=2\pi\nu$ represent the electric field amplitude, isotropic luminosity, and angular frequency of an FRB pulse, respectively. $\omega_B$ is the electron cyclotron frequency in a background magnetic field $B_{\text{bg}}$. In this paper, the background magnetic field strength is assumed to decrease as $r^{-3}$ in the magnetosphere. Here, $r$ is the distance from the magnetar, whose surface magnetic field strength and radius are set as $B_{\text{s}}=\left(10^{15} \text{ G}\right)B_{\text{s,15}}$ and $R_{\text{s}}=10^6\text{ cm}$, respectively. Although the exact scattering cross section of an electron within an FRB inside the magnetosphere cannot be derived, some approximate expressions are available within specific parameter space, as shown in Figure \ref{scacrosec}. Scattering cross sections of an electron in an X-mode FRB in different regions are summarized as follows.

If an FRB is emitted deep within the magnetosphere (e.g., $r_{\text{em}}\sim10 R_{\text{s}}\sim 10^7\text{ cm}$), then near the emission site, the electric field of the FRB  wave is significantly weaker than the background magnetic field, i.e.,
\begin{equation}
	\frac{E_0}{B_{\text{bg}}}=\frac{a_0}{\omega_B/\omega}\approx 8.2\times10^{-4}L_{\text{iso,42}}^{1/2}B_{\text{s,15}}^{-1}r_7^2.
\end{equation}
In this case, the electron motion is dominated by the strong background magnetic field. The scattering cross section of an electron is \citep{Canuto1971}
\begin{equation}
	\sigma_{\text{sc}}\approx\left(\frac{\omega_B}{\omega}\right)^{-2}\sigma_T\approx\left(1.3\times10^{-19}\nu_9^2 B_{\text{s,15}}^{-2} r_7^6\right) \sigma_T.
\end{equation}

According to Equation (\ref{a0omegab}), the background magnetic field decays faster than the electric field of an FRB as the distance increases. Therefore, a transition radius 
\begin{equation}
	r_{\text{tra}}\approx\left(3.5\times10^8\text{ cm}\right)L_{\text{iso,42}}^{-1/4}B_{\text{s,15}}^{1/2}
\end{equation}
can be reached at where $a_0=\omega_B/\omega$. Beyond this radius, the electric field strength of the FRB exceeds the background magnetic field. \cite{Beloborodov2022} discovered that in the region where $1<\omega_B/\omega<a_0$ satisfies, electrons in an electromagnetic wave exhibit an interesting behavior. They interact with the wave through frequent resonance events. The radiation reaction limit can be reached when the equilibrium between electron acceleration and cooling is established. The electron's Lorentz factor thus can be approximately given by
\begin{equation}
	\begin{aligned}
		\gamma_{\text{rrl}}&\sim\left(\frac{c}{r_e a_0 \omega}\right)^{3/8}\left(\frac{\omega_B}{\omega}\right)^{1/4}\\
		&\approx\left(1.5\times10^4\right) \nu_9^{-1/4}L_{\text{iso,42}}^{-3/16}B_{s,15}^{1/4}r_9^{-3/8}.
	\end{aligned}
	\label{gammarrl}
\end{equation}
The time to reach the equilibrium is
\begin{equation}
	\begin{aligned}
		t_{\text{rrl}}&\sim\frac{1}{a_0^2 \omega }\left(\frac{c}{r_e a_0 \omega}\right)^{7/8}\left(\frac{\omega_B}{\omega}\right)^{1/4}\\
		&\approx \left(1.3\times10^{-7}\text{ ms}\right)\nu_9^{3/4}L_{\text{iso,42}}^{-23/16}B_{\text{s,15}}^{1/4}r_{9}^{17/8},
		\label{trrl}
	\end{aligned}
\end{equation}
which is much shorter than the typical millisecond duration of FRBs. The scattering cross section then can be estimated as 
\begin{equation}
	\sigma_{\text{sc}}\sim\gamma_{\text{rrl}}^2 \sigma_T\approx \left(2.4\times10^8 \nu_9^{-1/2}L_{\text{iso,42}}^{-3/8}B_{\text{s,15}}^{1/2}r_{9}^{-3/4}\right)\sigma_T.
	\label{sigmarrl}
\end{equation}

In the region $\omega_B/\omega<1<a_0$, the scattering cross section is \citep{Beloborodov2022}
\begin{equation}
	\sigma_{\text{sc}}\sim a_0^2 \sigma_T\approx\left(5.3\times10^{4}\nu_9^{-2}L_{\text{iso,42}}r_{11}^{-2}\right)\sigma_T.
\end{equation}
One can see that the cross section in the region $1<\omega_B/\omega<a_0$ is significantly larger than that in other regions. Furthermore, it should be noted that these expressions for the scattering cross section are valid when the propagation direction of the FRB is perpendicular to the background magnetic field.

\subsection{Two-photon Annihilation Reaction}\label{tpannihilation}

Besides the enhancement of scattering cross sections, another important factor that likely blocks FRBs is that electron--positron pair plasma accelerated by the FRB emits gamma photons through curvature radiation. These photons further generate additional secondary electron--positron pairs through the two-photon annihilation process $\gamma+\gamma\to e^{-}+e^{+}$, which greatly increases the plasma number density.

The characteristic frequency of curvature radiation photons is \citep{Rybicki1991}
\begin{equation}
	\omega_c\approx\frac{3}{2}\frac{\gamma^3 c}{r_c},
\end{equation}
where
\begin{equation}
	r_c\approx \left(\frac{2ce^2 \gamma^4}{3P_e}\right)^{1/2}\approx\frac{\sqrt{2}c\gamma}{a_0 \omega}
	\label{curradiacur}
\end{equation}
is the curvature radius of the electron trajectory. The averaged emission power is
\begin{equation}
	\left<P_e\right>=\left<S\right>\sigma_{\text{sc}},
	\label{emipowperpen}
\end{equation}
where $\left<S\right>=mca_0^2\omega^2/8\pi r_e$ is the averaged incident flux of the FRB pulse. One can evaluate the ratio between the energy carried by an emitted photon and the electron rest energy,
\begin{equation}
	\begin{aligned}
		\frac{\hbar \omega_c}{m c^2}&\approx\frac{3\hbar}{2\sqrt{2}mc^2}a_0 \omega \gamma^2\\
		&\sim\left\{
		\begin{array}{ll}
			46.4 \nu_9^{-1/2}L_{\text{iso,42}}^{1/8}B_{\text{s,15}}^{1/2}r_{9}^{-7/4}, & \gamma=\gamma_{\text{rrl}} \\
			1.0\times10^{-4}\nu_9^{-2}L_{\text{iso,42}}^{3/2} r_{11}^{-3}, & \gamma=a_0
		\end{array}\right..
	\end{aligned}\label{homegacmcs}
\end{equation}
As a result, the characteristic photon energy in region $1<\omega_B/\omega<a_0$ is sufficient to create electron--positron pairs.

\begin{figure}
	\begin{center}
		\subfigure{
			\includegraphics[width=0.45\textwidth]{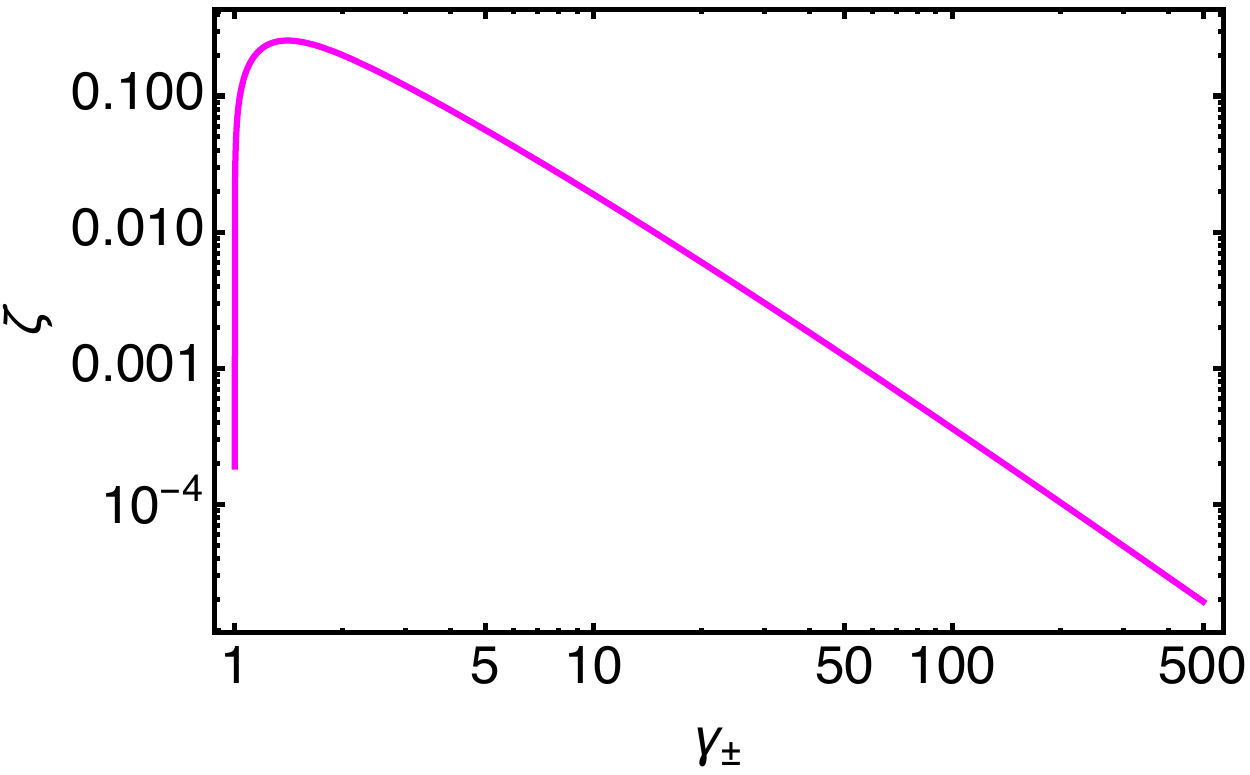}}
		\caption{The value of $\zeta$ as a function of $\gamma_{\pm}$.}
		\label{zetagamma}
	\end{center}
\end{figure}

An order of magnitude estimate on the number of pairs produced by these gamma photons is as follows. We assume that a test electron remains in the FRB pulse for a duration $T$, then its total emitted energy is approximately $P_e T$. If each emitted photon carries the characteristic energy $\hbar \omega_c$, then the number of photons emitted by the electron is
\begin{equation}
	N_{\gamma}\sim\frac{P_e T}{\hbar \omega_c}\approx\frac{2\sqrt{2}r_e mc}{9\hbar}a_0 \omega T\approx3.3\times10^{8}L_{\text{iso,42}}^{1/2}r_9^{-1}T_{-3}.
\end{equation}
The number density of an initial pair plasma is normally described by \citep{Goldreich1969}
\begin{equation}
	n\approx\xi n_{\text{GJ}}\approx\left(7.0\times10^7 \text{ cm}^{-3}\right) \xi_3 B_{\text{s,15}}P^{-1} r_9^{-3},
\end{equation}
where $\xi$ is a multiplicity factor, $n_{\text{GJ}}$ is the Goldreich-Julian (GJ) density, and $P$ is the spin period of the magnetar, respectively. A simple estimate of the number density of gamma photons inside the FRB pulse should be
\begin{equation}
	n_{\gamma}\sim N_{\gamma} n \sim \left(2.3\times10^{16} \text{ cm}^{-3}\right)\xi_3 L_{\text{iso,42}}^{1/2}B_{\text{s,15}}P^{-1}r_9^{-4}T_{-3}.
\end{equation}
For approximately isotropically distributed photons in the FRB pulse, the number of pairs generated by the photons emitted by an electron through the reaction $\gamma+\gamma\to e^{-}+e^{+}$ can be further estimated as
\begin{equation}
	\mathcal{M}\sim N_{\gamma}\overline{\sigma}_{\gamma\gamma}cTn_{\gamma}\sim 1.5\times10^{5}\xi_3 \zeta_{-3} L_{\text{iso,42}}B_{\text{s,15}}P^{-1}r_9^{-5}T_{-3}^3,
	\label{multifac}
\end{equation}
where $\overline{\sigma}_{\gamma\gamma}=\zeta\sigma_T$ represents the mean scattering cross section of $\gamma\gamma$ collisions. The value of $\zeta$ in the center-of-momentum frame is given by \citep{Zhang2018a}
\begin{equation}
	\begin{aligned}
		\zeta=&\frac{3}{8}\gamma_{\pm}^{-2}\left[\left(-2-2\gamma_{\pm}^{-2}+\gamma_{\pm}^{-4}\right)\ln{\left(\gamma_{\pm}-\sqrt{\gamma_{\pm}^2 -1}\right)}\right.\\
		&\left.-\left(1+\gamma_{\pm}^{-2}\right)\sqrt{1-\gamma_{\pm}^{-2}}\right],
	\end{aligned}
\end{equation} 
where $\gamma_{\pm}=\hbar\omega_c/\left(mc^2\right)$ is the Lorentz factor of outgoing pairs, given by Equation (\ref{homegacmcs}). The value of $\zeta$ as a function of $\gamma_\pm$ is plotted in Figure \ref{zetagamma}. The above calculation demonstrates that the large number of pair creation can significantly enhance the opacity of an FRB within the magnetosphere. More importantly, these new secondary electron--positron pairs can generate more pairs in a similar manner, and trigger a cascade process ($n\sim e^\mathcal{M}$), until most of the energy of the FRB is depleted \citep{Beloborodov2021}.

\section{Revisiting Particle Motion in an FRB Pulse}\label{frbpp}

The calculations in the previous section assume that the propagation direction of the FRB is perpendicular to the background magnetic field line. In fact, an FRB is more likely to propagate obliquely relative to the field line. In this section, we will generalize above results to the case of oblique propagation.

\begin{figure*}
	\begin{center}
		\subfigure{
			\includegraphics[width=0.7\textwidth]{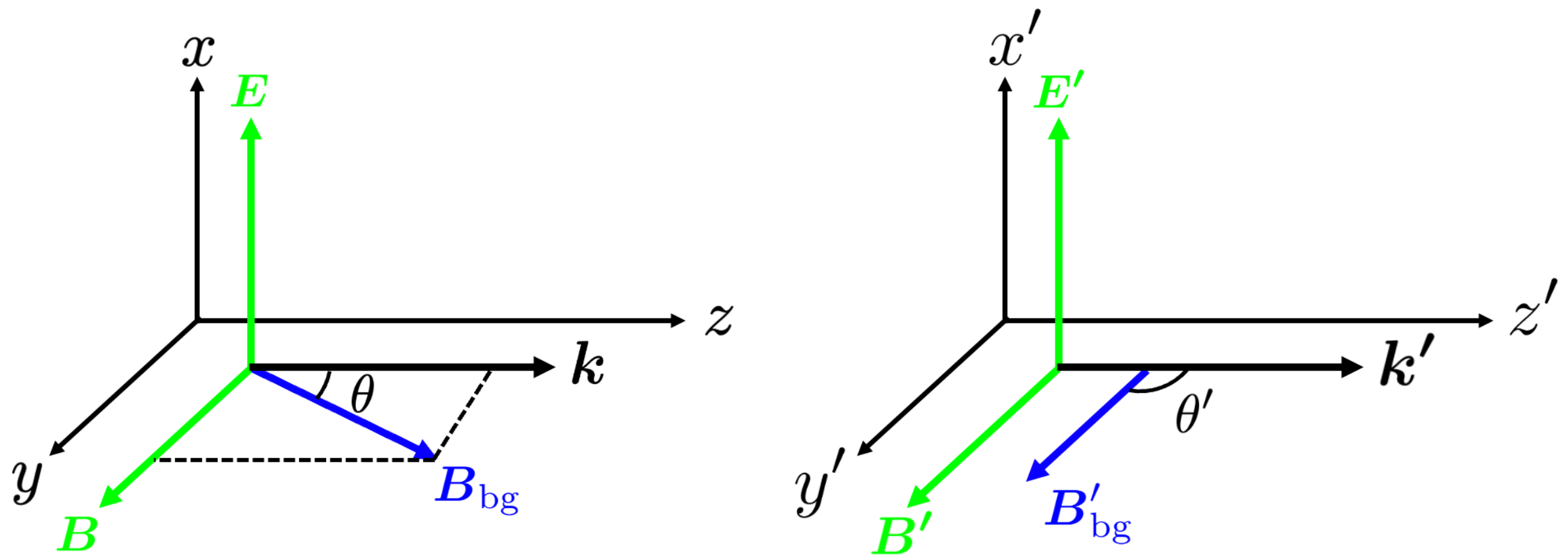}}
		\caption{A coordinate system established in the lab frame (left panel). An FRB is propagating along the $z-$ axis. The electric field and magnetic field of the FRB are along the $x-$ and $y-$ axes, respectively. The background magnetic field is in the $y-z$ plane. The angle between the wave vector and the background magnetic field is $\theta$. A coordinate system established in the frame $K^\prime$ (right panel), which is defined as a frame that slides along the background magnetic field line with a Lorentz factor $\gamma_f=1/\sin{\theta}$ in the lab frame. The FRB propagation direction is assumed to be along the $z^\prime-$ axis. The electric field and magnetic field are along the $x^\prime-$ and $y^\prime -$ axes, respectively. In the frame $K^\prime$, the background magnetic field has the same direction as the magnetic field of the FRB.}
		\label{coordsys}
	\end{center}
\end{figure*}

We first establish a coordinate system in the lab frame, as shown in the left panel of Figure \ref{coordsys}. We assume an FRB is propagating along the $z-$ axis. The electric field and magnetic field of the FRB are along the $x-$ and $y-$ axes, respectively. The background magnetic field is assumed to be in the $y-z$ plane, and has an angle $\theta$ with respect to the $z-$ axis. One can also define a reference frame $K^\prime$ that is sliding along the background magnetic field line with a Lorentz factor $\gamma_f=1/\sin{\theta}$. An advantage in such a reference frame is that the magnetic field direction of the FRB is aligned with the background magnetic field. A corresponding coordinate system thus can be established in this frame, as shown in the right panel of Figure \ref{coordsys} (we use a prime symbol to denote the quantity in the frame $K^\prime$). The relativistic transformations between the frame $K^\prime$ and the lab frame are \citep{Huang2024}
\begin{equation}
	a_0^\prime=a_0,\text{ }\frac{\omega_B^\prime}{\omega^\prime}=\frac{\omega_B}{\omega}\gamma_f,\text{ }\omega^\prime=\frac{\omega}{\gamma_f}.
	\label{retransf}
\end{equation}

\begin{figure*}
	\begin{center}
		\subfigure{
			\includegraphics[width=0.32\textwidth]{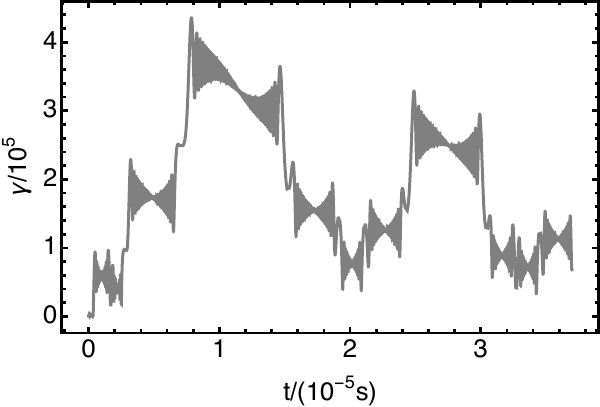}}
		\subfigure{
			\includegraphics[width=0.32\textwidth]{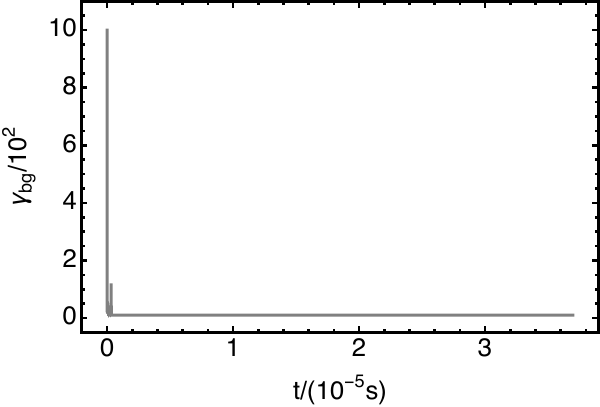}}
		\subfigure{
			\includegraphics[width=0.32\textwidth]{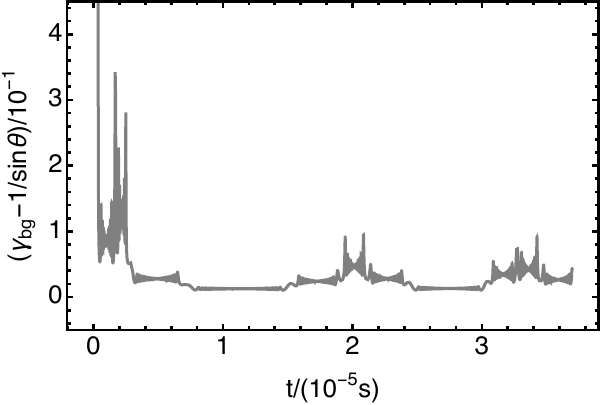}}
		\subfigure{
			\includegraphics[width=0.32\textwidth]{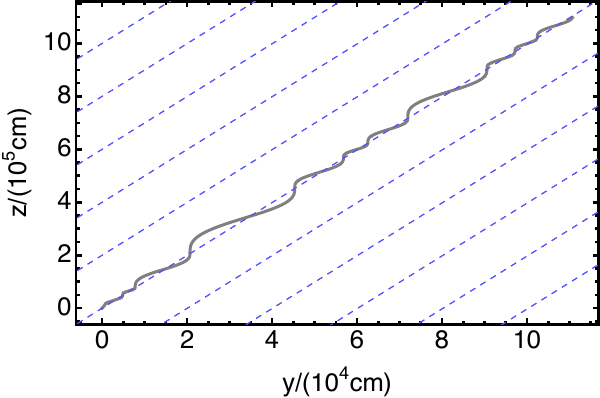}}
		\subfigure{
			\includegraphics[width=0.32\textwidth]{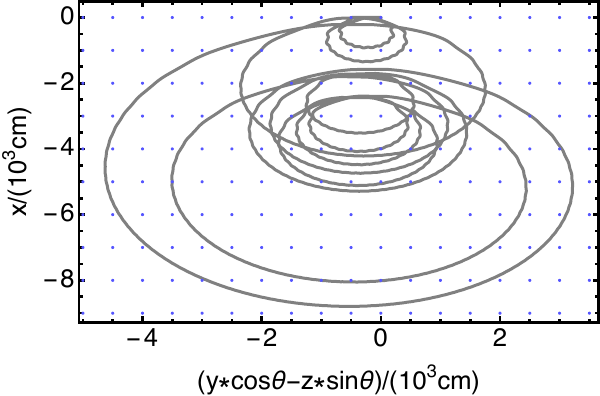}}
		\subfigure{
			\includegraphics[width=0.32\textwidth]{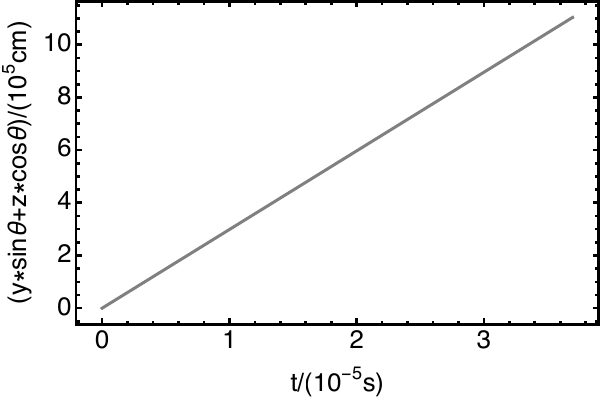}}
		\caption{The motion of a test particle in an FRB. The Lorentz factor of the particle as a function of time (upper left panel). The Lorentz factor of the velocity along the field line, $\gamma_{\text{bg}}\equiv\left(1-\beta_{\text{bg}}^2\right)^{-1/2}$ as a function of time (upper middle panel) and its zoom-in view (upper right panel). The trajectory of the particle in the $y-z$ plane (lower left panel) and in the plane perpendicular to the field line (lower middle panel). The blue dashed lines and dots indicate the direction of the field line. The distance that the particle travels along the field line with respect to time (lower right panel). Before entering the FRB, the particle has an initial Lorentz factor $\gamma_i=1000$ along the field line. As time goes by, the Lorentz factor corresponding to the velocity along the field line drops rapidly, and finally reaches a stable value $\gamma_{\text{bg}}\approx1/\sin{\theta}$. The parameters $a_0=2000$ and $\omega_B/\omega=40$ are adopted, which is equivalent to an FRB with isotropic luminosity $L_{\text{iso}}\approx1.3\times10^{41}\text{ erg s}^{-1}$ and distance from the magnetar $r\approx4.1\times10^9\text{ cm}$. The other adopted parameters are: $\theta=10^{-1}\text{ rad}$, $\nu=10^9\text{ Hz}$, and $B_s=10^{15}\text{ G}$.}
		\label{labmotion}
	\end{center}
\end{figure*}

\begin{figure*}
	\begin{center}
		\subfigure{
			\includegraphics[width=0.32\textwidth]{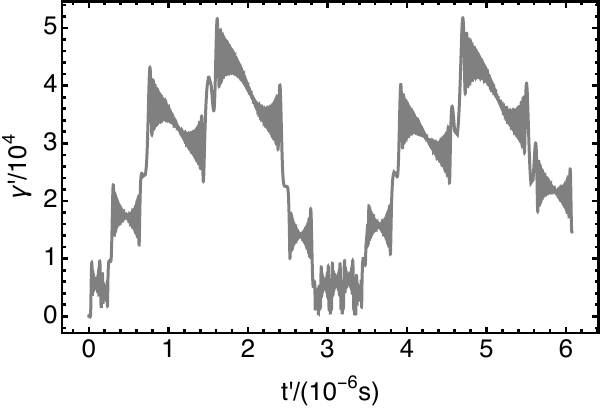}}
		\subfigure{
			\includegraphics[width=0.32\textwidth]{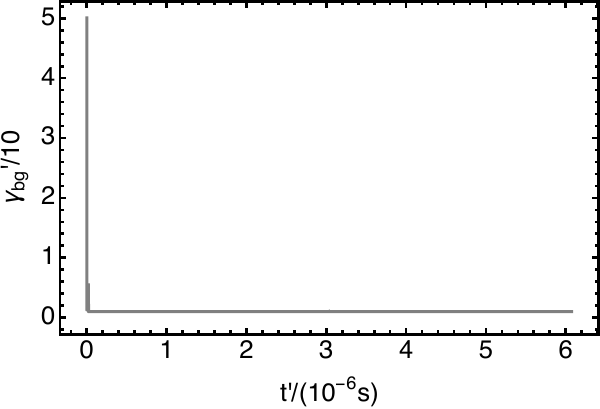}}
		\subfigure{
			\includegraphics[width=0.32\textwidth]{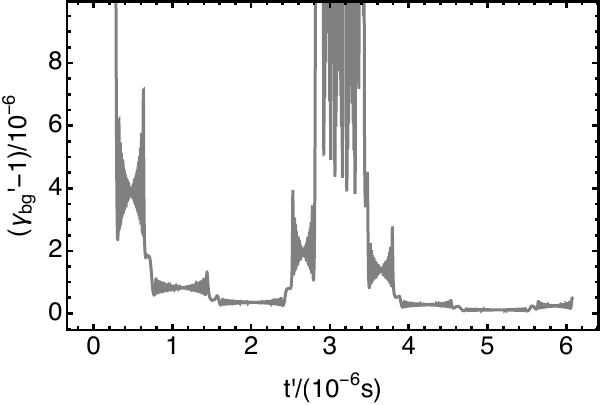}}
		\subfigure{
			\includegraphics[width=0.32\textwidth]{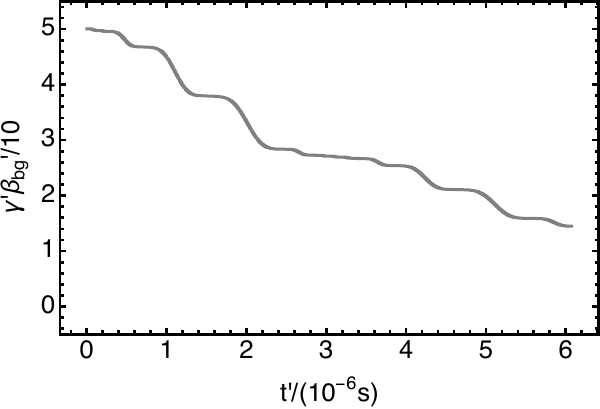}}
		\subfigure{
			\includegraphics[width=0.32\textwidth]{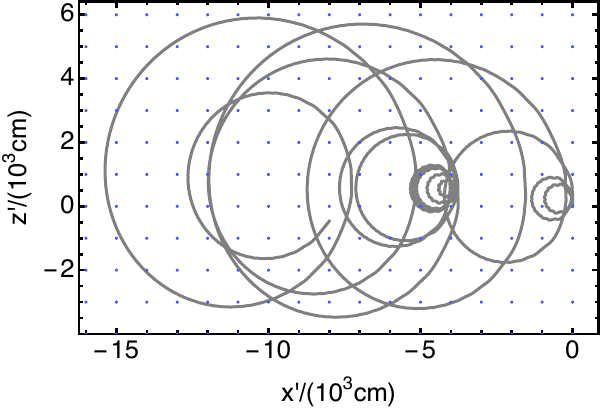}}
		\subfigure{
			\includegraphics[width=0.32\textwidth]{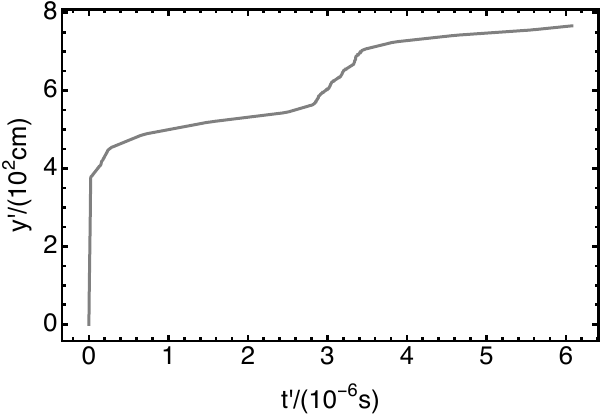}}
		\caption{The motion of a test particle in an FRB observed in the frame $K^\prime$. The Lorentz factor of the particle as a function of time (upper left panel). The time evolution of the Lorentz factor of the velocity along the field line (upper middle panel) and its zoom-in view (upper right panel). The dimensionless momentum of the particle along the field line as a function of time (lower left panel). The trajectory of the particle in the $x^\prime-z^\prime$ plane (lower middle panel). The blue dots indicate the direction of the field line. The distance that the particle travels along the $y^\prime-$ axis with respect to time (lower right panel). At the beginning, the particle has an initial Lorentz factor $\gamma_i^\prime\approx50$ (corresponding to $\gamma_i=1000$ in the lab frame) along the field line. As time goes by, the velocity of the particle along the field line drops rapidly, and then nearly vanishes. The adopted parameters are: $a_0^\prime=2000$, $\omega_B^\prime/\omega^\prime\approx400$, $\theta^\prime=\pi/2 \text{ rad}$, and $\nu^\prime=6.3\times10^8\text{ Hz}$, which are transformed from those parameters adopted in Figure \ref{labmotion} from relativistic transformations (\ref{retransf}).}
		\label{Kprimemotion}
	\end{center}
\end{figure*}

An example for the motion of a test particle in the lab frame is shown in Figure \ref{labmotion}. The motion of the particle has three interesting properties. Firstly, the Lorentz factor $\gamma$ is boosted to a high value by continuous resonance events. Secondly, the velocity of the particle along the field line $\beta_{\text{bg}}$ quickly drops down and finally becomes nearly constant. Thirdly, the particle exhibits significant gyration around the field line. One can also simulate the motion of the particle independently in the frame $K^\prime$ by adopting the same parameters as those in Figure \ref{labmotion} from relativistic transformations (\ref{retransf}). The motion of the particle in the frame $K^\prime$ is shown in Figure \ref{Kprimemotion}. Although the particle possesses a high Lorentz factor $\gamma^\prime$ and also exhibits significant gyration around the field line, its velocity component along the field line nearly vanishes. This is compatible with that observed in the lab frame, i.e., the particle nearly slides along the field line at the same velocity as the reference frame $K^\prime$,
\begin{equation}
	\beta_{\text{bg}}\approx\beta_f\equiv\left(1-1/\gamma_f^2\right)^{1/2}=\cos{\theta}.
	\label{betabg}
\end{equation}

This phenomenon can be explained as follows. Let us assume that the particle possesses an initial Lorentz factor $\gamma_i\gg 1$ along the field line in the lab frame.\footnote{In fact, $\gamma_i\gg1$ is controversial \citep{Beloborodov2023}. However, we show here that even if the initial Lorentz factor $\gamma_i\gg1$, it cannot effectively influence the motion of a particle in an FRB.} The Lorentz factor $\gamma_i$ corresponds to an initial velocity 
\begin{equation}
	\beta_i=\left(1-1/\gamma_i^2\right)^{1/2}\approx1-\frac{1}{2\gamma_i^2}.
\end{equation}
This velocity observed in the frame $K^\prime$ can be derived by a relativistic transformation,
\begin{equation}
	\beta_i^\prime=\frac{\beta_i-\beta_f}{1-\beta_i \beta_f}\approx1-\frac{1+\cos{\theta}}{\cos{\theta}+2\gamma_i^2 \left(1-\cos{\theta}\right)},
\end{equation}
which corresponds to a Lorentz factor
\begin{equation}
	\gamma_i^\prime=\left(1-\beta_i^{\prime 2}\right)^{-1/2}\approx\gamma_i\left(\frac{1-\cos{\theta}}{1+\cos{\theta}}\right)^{1/2},
\end{equation}
where in the second equality, we have assumed $\gamma_i^2\left(1-\cos{\theta}\right)\gg1/2$. When $\theta\ll1$, one has $\gamma_i^\prime\approx \gamma_i \theta/2$. This indicates that any large initial Lorentz factor in the lab frame will be reduced by a factor $\theta/2$ in the frame $K^\prime$. For example, when $\gamma_i=1000$ and $\theta=10^{-1}\text{ rad}$, the initial Lorentz factor $\gamma_i^\prime\approx50$, as shown in the upper middle panel of Figure \ref{Kprimemotion}.

In the reference frame $K^\prime$, along the direction of the field line, the particle only experiences a radiation reaction force, which is anti-parallel to the direction of its velocity. Therefore, the momentum component of the particle along the field line loses continuously, as shown in the lower left panel of Figure \ref{Kprimemotion}. One then has
\begin{equation}
	\gamma^\prime \beta_{\text{bg}}^\prime<\gamma_i^\prime \beta_{i}^\prime\Longrightarrow\beta_{\text{bg}}^\prime<\frac{\gamma_i^\prime}{\gamma^\prime}\beta_{i}^\prime\ll 1,
	\label{momentumbg}
\end{equation}
where we have used Equation (\ref{gammarrl}) but in the frame $K^\prime$,
\begin{equation}
	\gamma^\prime\sim \left(\frac{c}{r_e a_0^\prime \omega^\prime}\right)^{3/8}\left(\frac{\omega_B^\prime}{\omega^\prime}\right)^{1/4}=\gamma_{\text{rrl}}\gamma_f^{5/8}\gg \gamma_i^\prime.
	\label{gammaprime}
\end{equation}
In conclusion, a large initial Lorentz factor ($\gamma_i\sim1000$) in the lab frame is not important to influence the motion of the particle when $\theta\ll1$. Once the particle is exposed in the FRB pulse, its velocity along the background magnetic field line quickly synchronizes with the velocity of the frame $K^\prime$, which is given by Equation (\ref{betabg}).

The Lorentz factor $\gamma^\prime$ in the frame $K^\prime$ corresponds to a velocity $\beta^\prime\approx\left(1-1/\gamma^{\prime 2}\right)^{1/2}$, which is almost perpendicular to the field line. In the lab frame, the velocity perpendicular to the field line is obtained by a relativistic transformation, $\beta_{\text{bg},\perp}\approx \beta^\prime/\gamma_f$. Therefore, the Lorentz factor of the particle in the lab frame is
\begin{equation}
	\widetilde{\gamma}=\left(1-\beta_{\text{bg},\perp}^2-\beta_{\text{bg}}^2\right)^{-1/2}\approx \gamma^\prime \gamma_f=\gamma_{\text{rrl}}\gamma_f^{13/8},
\end{equation}
where we use the tilde symbol to denote the generalization of the quantity in the case of oblique propagation. Although the Lorentz factor is boosted by a factor $\gamma_f^{13/8}$ compared to the perpendicular case, the emission power of the particle is reduced. A general expression of the total emission power of a particle oscillating in an FRB is \citep{Beloborodov2022,Huang2024}
\begin{equation}
	\widetilde{P}_e\sim\frac{1}{3}\frac{e^2 a_0^2 \omega^2}{c}\widetilde{\gamma}^2 \left(1-\beta_z\right)^2\sim\frac{1}{3}\frac{e^2 a_0^2 \omega^2}{c}\gamma_{\text{rrl}}^2 \gamma_f^{-3/4},
\end{equation}
where we have used $\beta_z\approx\beta_{\text{bg}}\cos{\theta}\approx\cos^2{\theta}$.
The emission power reduces to Equation (\ref{emipowperpen}) when $\theta=\pi/2$. The scattering cross section of an electron in an FRB thus is
\begin{equation}
	\widetilde{\sigma}_{\text{sc}}=\frac{\widetilde{P}_e}{\left<S\right>}\sim \sigma_T \gamma_{\text{rrl}}^2 \gamma_f^{-3/4}.
\end{equation}
For the case of oblique propagation, one can see that the scattering cross section is reduced by a factor $\gamma_f^{-3/4}$.

The generalized characteristic frequency of the curvature radiation photon is
\begin{equation}
	\widetilde{\omega}_c\approx\frac{3}{2}\frac{\widetilde{\gamma}^3 c}{\widetilde{r}_c},
	\label{omegactilde}
\end{equation}
where $\widetilde{r}_c$ is the generalized curvature radius, which can be estimated as follows. In the frame $K^\prime$, the curvature radius is
\begin{equation}
	\begin{aligned}
		r_c^\prime&\approx \frac{\sqrt{2}c\gamma^\prime}{a_0^\prime\omega^\prime}\approx\frac{\sqrt{2}c\gamma_{\text{rrl}}}{a_0 \omega}\gamma_f^{13/8}\\
		&\approx\left(4.5\text{ cm}\right)\nu_9^{-1/4}L_{\text{iso,42}}^{-11/16}B_{s,15}^{1/4}r_9^{5/8}\gamma_f^{13/8}.
	\end{aligned}
\end{equation}
It is worthy of noting that the transverse length remains invariant under the relativistic transformation, i.e., $r_{\text{c},\text{bg},\perp}\approx r_c^\prime$. Within a time interval $2\pi r_{\text{c},\text{bg},\perp}/\left(c\beta_{\text{bg},\perp} \right)$, the distance that the particle travels along the field line is
\begin{equation}
	r_{c,\text{bg}}\approx\frac{2\pi r_{c,\text{bg},\perp}}{c\beta_{\text{bg},\perp}}c\beta_{\text{bg}}\approx 2\pi r_c^\prime \cot{\theta}.
\end{equation}
The curvature radius of the particle's trajectory thus can be estimated by
\begin{equation}
	\begin{aligned}
		\widetilde{r}_c&\sim\left(r_{c,\text{bg}}^2+r_{c,\text{bg},\perp}^2\right)^{1/2}\approx\left(4\pi^2 \cot^2{\theta}+1\right)^{1/2}r_c^\prime\\
		&\approx\frac{\sqrt{2}c\gamma_{\text{rrl}}}{a_0 \omega}\left\{
		\begin{array}{ll}
			2\pi \gamma_f^{21/8}, & \theta\lesssim81^\circ\\
			1, & \theta\gtrsim 81^\circ
		\end{array}\right..
	\end{aligned}
\end{equation}
A propagation angle $\theta<81^\circ$ is more reasonable, and will be adopted in the following discussion. For $\nu=10^9\text{ Hz}$, $a_0=2000$, $\omega_B/\omega=40$, and $\theta=0.1\text{ rad}$, the curvature radius $\widetilde{r}_c\approx1.2\times10^5\text{ cm}$, which is consistent with that shown in the lower left panel of Figure \ref{labmotion}. One can recalculate the ratio between the characteristic energy of the emitted photon and the electron rest energy,
\begin{equation}
	\begin{aligned}
		\frac{\hbar \widetilde{\omega}_c}{mc^2}&\approx\frac{3\hbar}{4\sqrt{2}\pi m c^2}a_0 \omega \gamma_{\text{rrl}}^2 \gamma_f^{9/4}\\
		&\approx7.4 \nu_9^{-1/2}L_{\text{iso,42}}^{1/8}B_{s,15}^{1/2}r_9^{-7/4}\gamma_f^{9/4}.
	\end{aligned}
\end{equation}
Outside the light cylinder $R_{\text{LC}}\approx\left(4.8\times10^9\text{ cm}\right)P_0$, the toroidal field gradually dominates the magnetic field configuration. Therefore, the characteristic energy of emitted photons is insufficient to effectively generate electron--positron pairs. We will focus on the scattering problem inside the magnetosphere.

\section{Reduce Scattering via Various Approaches}\label{approaches}

\begin{figure}
	\begin{center}
		\subfigure{
			\includegraphics[width=0.3\textwidth]{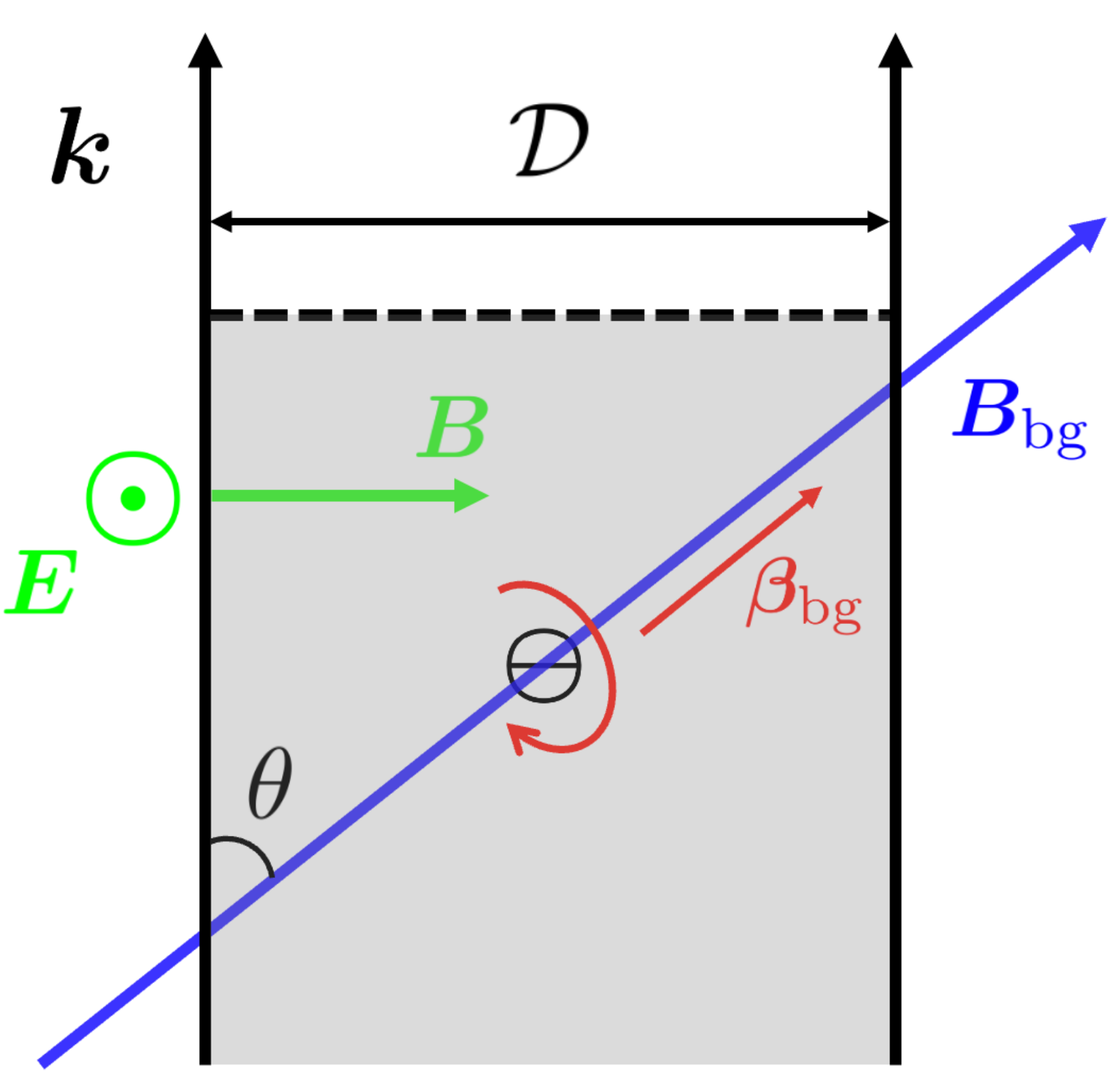}}
		\caption{A schematic illustration of the motion of an electron in an X-mode FRB pulse in the lab frame. The pulse (gray) propagates at an angle $\theta$ relative to the background magnetic field (blue). The directions of the electric and magnetic fields of the pulse are marked with green arrows. The transverse spatial size of the pulse is denoted as $\mathcal{D}$. In addition to the cyclotron motion, the electron is nearly sliding along the magnetic field line with a constant velocity $\beta_{\text{bg}}$.}
		\label{bandk}
	\end{center}
\end{figure}

\begin{figure*}
	\begin{center}
		\subfigure{
			\includegraphics[width=0.32\textwidth]{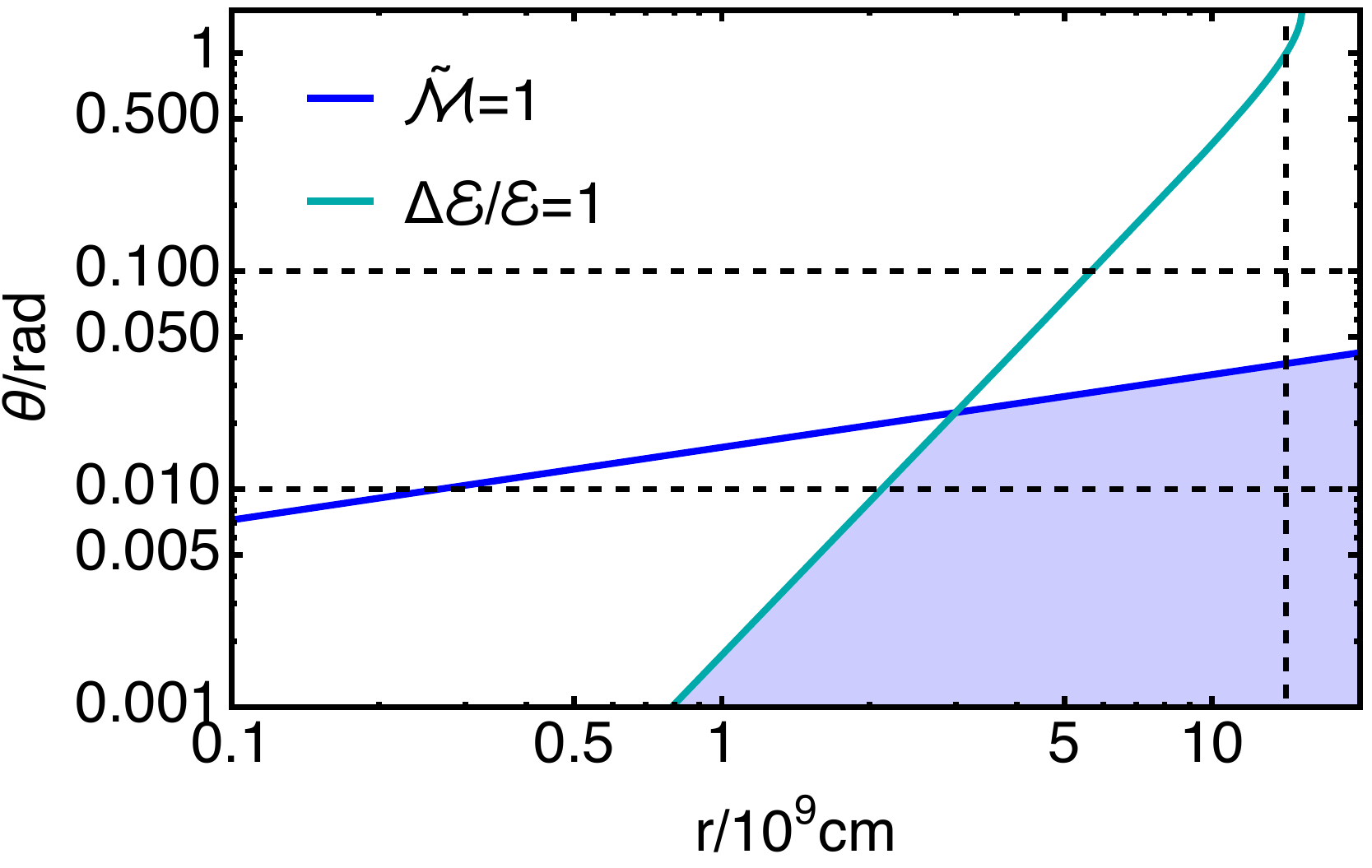}}
		\subfigure{
			\includegraphics[width=0.32\textwidth]{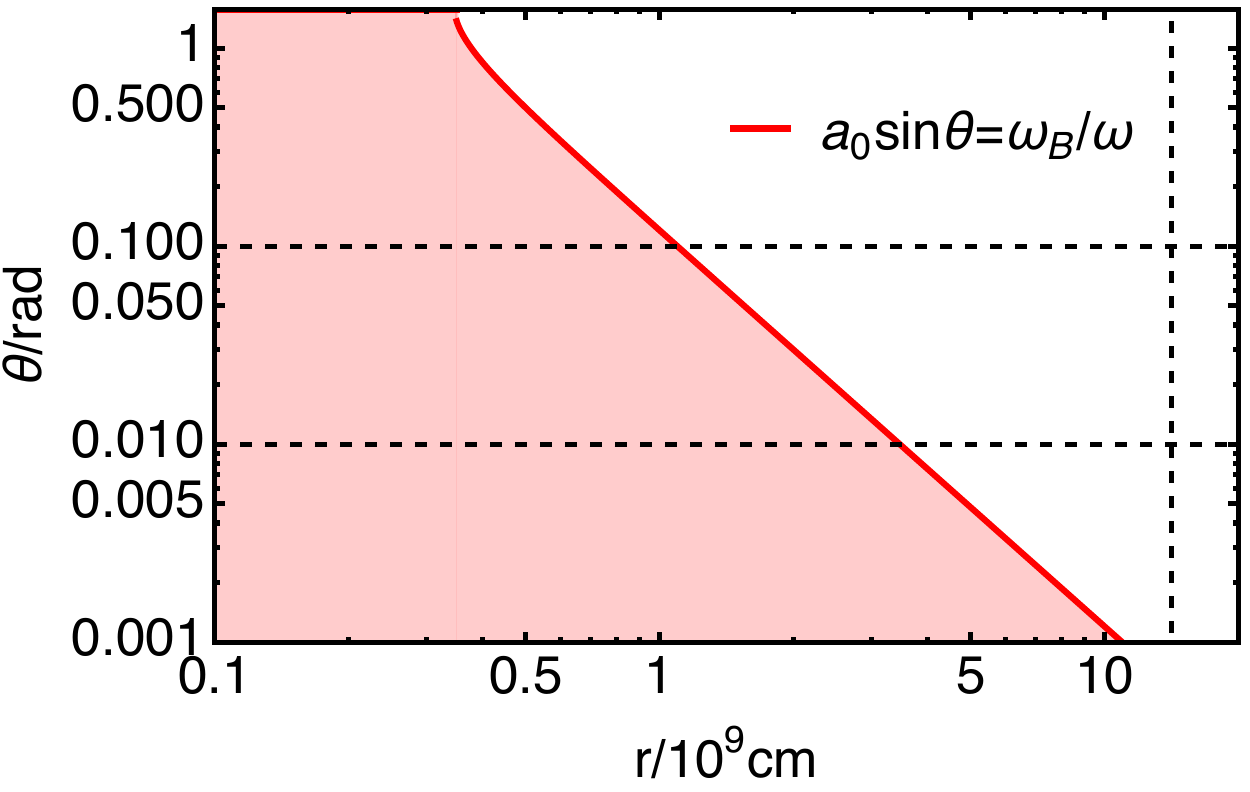}}
		\subfigure{
			\includegraphics[width=0.32\textwidth]{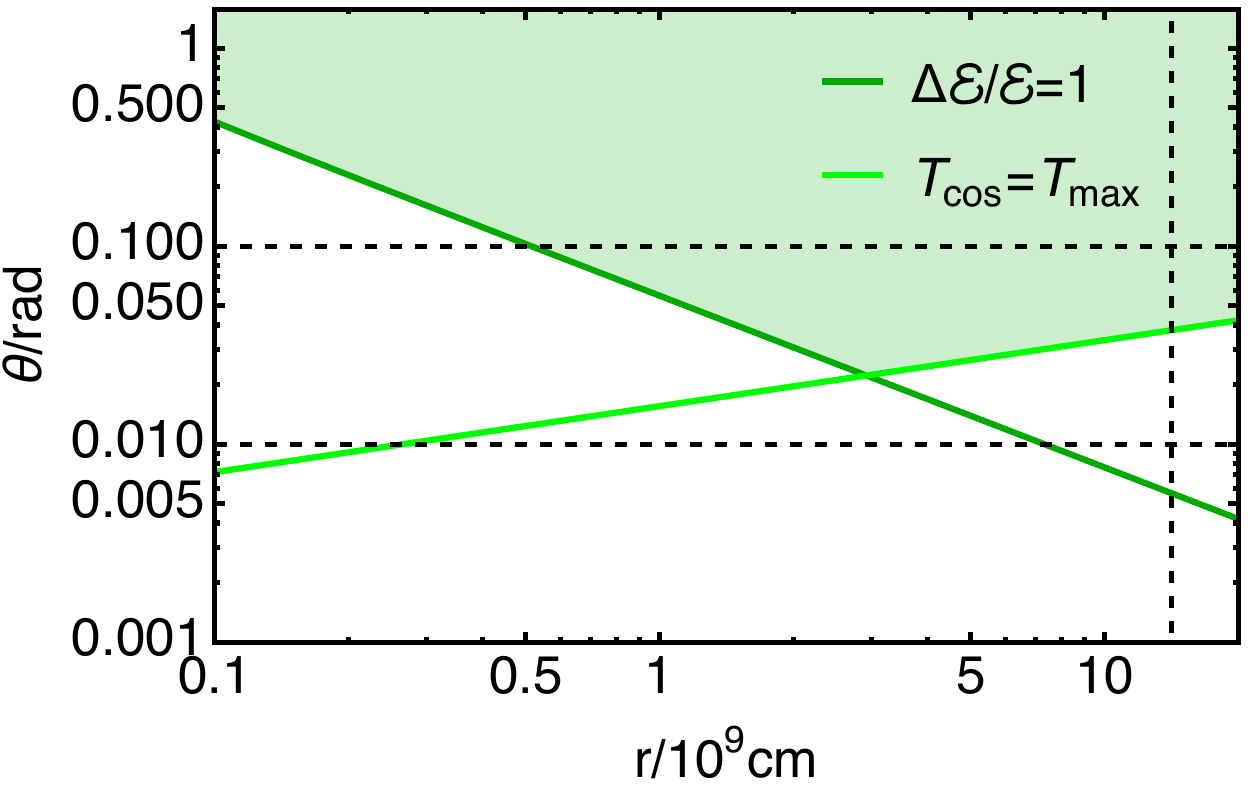}}
		\caption{The constraints on the value of $\theta$ as a function of distance $r$ in various approaches. The physically allowed regions in different approaches are highlighted by different colors. The vertical dashed line represents the light cylinder radius. The horizontal dashed lines represent $\theta=10^{-2}$ and $10^{-1}\text{ rad}$. The adopted parameters are: $\xi=10^3$, $\zeta=10^{-3}$, $\nu=10^9\text{ Hz}$, $L_{\text{iso}}=10^{42}\text{ erg s}^{-1}$, $B_s=10^{15}\text{ G}$, $P=2.95\text{ s}$, and $\mathcal{D}=\mathcal{D}_{\text{cr}}$.}
		\label{thetacrr}
	\end{center}
\end{figure*}

Based on above calculations, Equation (\ref{multifac}) ultimately can be rewritten as
\begin{equation}
	\widetilde{\mathcal{M}}\sim6.1\times 10^6 \xi_3 \zeta_{-3} L_{\text{iso,42}}B_{s,15}P^{-1}r_9^{-5}T_{-3}^3 \gamma_f^{-6}.
	\label{tildemvalue}
\end{equation}
One may argue that for a small propagation angle such as $\theta\sim0.1\text{ rad}$, $\gamma_f^{-6}\sim10^{-6}$ is enough to cancel out the large number of pair creation. However, if the angle $\theta$ is small, the time $T$ that the particle stays in the wave can also deviate from one millisecond significantly. This can be seen more clearly in Figure \ref{bandk}. In the lab frame, the particle is nearly sliding along the background magnetic field with a constant velocity $\beta_{\text{bg}}\approx\cos{\theta}$. The velocity components parallel and perpendicular to the wave vector are $\beta_{\boldsymbol{k}}=\beta_{\text{bg}} \cos{\theta}\approx\cos^2{\theta}$ and $\beta_{\boldsymbol{k},\perp}=\beta_{\text{bg}} \sin{\theta}\approx\sin{\theta}\cos{\theta}$, respectively. In general, there are three available choices for $T$. The first choice is the time that a particle takes to penetrate the pulse along the wave vector,
\begin{equation}
	T_{\text{pen}}\approx\frac{\left(10^{-3}\text{ s}\right)}{1-\beta_{\boldsymbol{k}}}\approx\left(10^{-3}\text{ s}\right)\gamma_f^2,
	\label{tgammaf}
\end{equation}
where the duration of the pulse is set as one millisecond. In this case, replacing $T=T_{\text{pen}}$ in Equation (\ref{tildemvalue}) is incapable of reducing the value of $\widetilde{\mathcal{M}}$ effectively. Nevertheless, another possibility is that the theoretical maximum,
\begin{equation}
	T_{\text{max}}\approx \frac{r}{c}\approx\left(3.3\times10^{-2}\text{ s}\right)r_9
\end{equation}
is smaller than the value of $T_{\text{pen}}$. In this case, $T$ in Equation (\ref{tildemvalue}) should be replaced by $T_{\text{max}}$. On the one hand, in order to control pair creation ($\widetilde{\mathcal{M}}<1$), one requires
\begin{equation}
	\sin{\theta}< 1.3\times10^{-2} \xi_3^{-1/6} \zeta_{-3}^{-1/6} L_{\text{iso,42}}^{-1/6}B_{s,15}^{-1/6}P^{1/6}r_9^{1/3}.
	\label{confromm}
\end{equation}
On the other hand, the lost energy of the FRB due to scattering by particles is
\begin{equation}
	\Delta\mathcal{E}\sim \widetilde{P}_e T N_e\sim \left<S\right>\widetilde{\sigma}_{\text{sc}}T\xi n_{\text{GJ}} \theta_{j}^2 r^3,
\end{equation}
where $N_e$ is the number of particles scattering the FRB, $r$ is the radius where scattering is most severe, and $\theta_{j}$ is the radiation cone half-opening angle of the FRB pulse. The total realistic energy of the FRB is
\begin{equation}
	\mathcal{E}\sim\left(10^{-3}\text{ s}\right)\left<S\right> \theta_{j}^2 r^2.
\end{equation}
Therefore, the ratio of energy loss is
\begin{equation}
	\begin{aligned}
		\frac{\Delta\mathcal{E}}{\mathcal{E}}&\sim\xi n_{\text{GJ}} \widetilde{\sigma}_{\text{sc}}r T_{-3}\\
		&\sim3.5\times10^2 \xi_3 \nu_9^{-1/2}L_{\text{iso,42}}^{-3/8}B_{s,15}^{3/2}P^{-1}r_9^{-7/4}\gamma_f^{-3/4}.
	\end{aligned}
\end{equation}
The value of $\Delta\mathcal{E}/\mathcal{E}$ is not supposed to exceed unity, which implies
\begin{equation}
	\sin{\theta}<4.1\times10^{-4}\xi_3^{-4/3}\nu_9^{2/3}L_{\text{iso,42}}^{1/2}B_{s,15}^{-2}P^{4/3}r_9^{7/3}.
\end{equation}
The joint constraint on $\theta$ as a function of distance $r$ by combining both $\widetilde{\mathcal{M}}<1$ and $\Delta\mathcal{E}/\mathcal{E}<1$ is shown in the left panel of Figure \ref{thetacrr}.

Another approach to reduce scattering is as follows. As we have mentioned before, the strong scattering condition $1<\omega_B/\omega<a_0$ is valid only when $\theta=\pi/2$. Therefore, the strong scattering can occur only when $1<\omega_B^\prime/\omega^\prime<a_0^\prime$ is satisfied in the frame $K^\prime$. After the relativistic transformation (\ref{retransf}), this becomes $\sin{\theta}<\omega_B/\omega<a_0\sin{\theta}$. A critical propagation angle can be reached when
\begin{equation}
	a_0 \sin{\theta}\sim \omega_B/\omega,
	\label{a0omegabomega}
\end{equation}
which is identical to the empirical formula obtained by \cite{Qu2022} through numerical simulation. \footnote{Another explanation that gives the same critical propagation angle is presented in \cite{Qu2024}.} Therefore, another effective way to reduce scattering is to require $a_0 \sin{\theta}<\omega_B/\omega$, i.e.,
\begin{equation}
	\sin{\theta}<1.2\times10^{-1} L_{\text{iso,42}}^{-1/2}B_{s,15}r_9^{-2}.
	\label{confroma0omegab}
\end{equation}
The corresponding constraint on $\theta$ as a function of distance $r$ is shown in the middle panel of Figure \ref{thetacrr}. Combining the aforementioned results, one can draw a conclusion that a propagation angle $\theta\lesssim10^{-2}\text{ rad}$ along the propagation path to the light cylinder can reduce scattering effectively. However, such an extremely small propagation angle seems somewhat unnatural unless the FRB can tilt field lines to make them get aligned with the propagation direction \citep{Qu2022}. Nevertheless, whether such a phenomenon can occur is still in doubt \citep{Beloborodov2023}.

Here, we propose another simple strategy to control pair creation without requiring the angle is as small as $10^{-2}\text{ rad}$. As shown in Figure \ref{bandk}, the time needed for the particle to cross the FRB pulse laterally is
\begin{equation}
	T_{\text{cro}}=\frac{\mathcal{D}}{c\beta_{\boldsymbol{k},\perp}}\approx\frac{\mathcal{D}}{c\sin{\theta}\cos{\theta}}.
	\label{treduced}
\end{equation}
If $T_{\text{cro}}<T_{\text{max}}$, i.e., an FRB has a sufficiently small transverse spatial size $\mathcal{D}<\left(10^9\text{ cm}\right)r_9 \sin{\theta}\cos{\theta}$, then the time $T$ in Equation (\ref{tildemvalue}) should be replaced by $T_{\text{cro}}$.
In this case, $\widetilde{M}<1$ is equivalent to
\begin{equation}
	\begin{aligned}
		\mathcal{D}< \mathcal{D}_{\text{cr}}
		\approx &\left(1.6\times10^5\text{ cm}\right)\xi_3^{-1/3}\zeta_{-3}^{-1/3}\cot{\theta}\\
		&\times L_{\text{iso,42}}^{-1/3}B_{\text{s,15}}^{-1/3}P^{1/3}r_9^{5/3}.
	\end{aligned}\label{lperpmax}
\end{equation}
An FRB pulse with a transverse size smaller than the critical value $\mathcal{D}_{\text{cr}}$ can effectively reduce the production of electron--positron pairs. The ratio of energy loss is
\begin{equation}
	\begin{aligned}
		\frac{\Delta\mathcal{E}}{\mathcal{E}}&\sim\xi n_{\text{GJ}} \widetilde{\sigma}_{\text{sc}}r T_{-3}\sim5.6\times10^{-2}\xi_3^{2/3}\zeta_{-3}^{-1/3}\nu_9^{-1/2}\\
		&\times L_{\text{iso,42}}^{-17/24}B_{s,15}^{7/6}P^{-2/3}r_9^{-13/12}\gamma_f^{5/4}\left(\frac{\mathcal{D}}{\mathcal{D}_{\text{cr}}}\right).
	\end{aligned}
\end{equation}
Therefore, $\Delta\mathcal{E}/\mathcal{E}<1$ is equivalent to 
\begin{equation}
	\begin{aligned}
		\sin{\theta}&>0.1\xi_3^{8/15}\zeta_{-3}^{-4/15}\nu_9^{-2/5}L_{\text{iso,42}}^{-17/30}\\
		&\times B_{s,15}^{14/15} P^{-8/15}r_9^{-13/15}\left(\frac{\mathcal{D}}{\mathcal{D}_{\text{cr}}}\right)^{4/5}.
		\label{tcroe}
	\end{aligned}
\end{equation}
The premise of these calculations, $T_{\text{cro}}<T_{\text{max}}$ indicates
\begin{equation}
	\begin{aligned}
		\sin{\theta}&>1.3\times10^{-2} \xi_3^{-1/6} \zeta_{-3}^{-1/6}\\
		&\times L_{\text{iso,42}}^{-1/6}B_{s,15}^{-1/6}P^{1/6}r_9^{1/3}\left(\frac{\mathcal{D}}{\mathcal{D}_{\text{cr}}}\right)^{1/2},
		\label{tcrotmax}
	\end{aligned}
\end{equation}
which is opposite to Equation (\ref{confromm}) when $\mathcal{D}=\mathcal{D}_{\text{cr}}$. The constraint on $\theta$ combing Equations (\ref{tcroe}) and (\ref{tcrotmax}) is shown in the right panel of Figure \ref{thetacrr}. In summary, the valid regions of these three constraints almost cover the parameter space where scattering is most severe ($r\sim10^{9}-10^{10}\text{ cm}$), which indicates FRBs can indeed escape from magnetar magnetospheres through different approaches.

\begin{figure*}
	\begin{center}
		\subfigure{
			\includegraphics[width=0.4\textwidth]{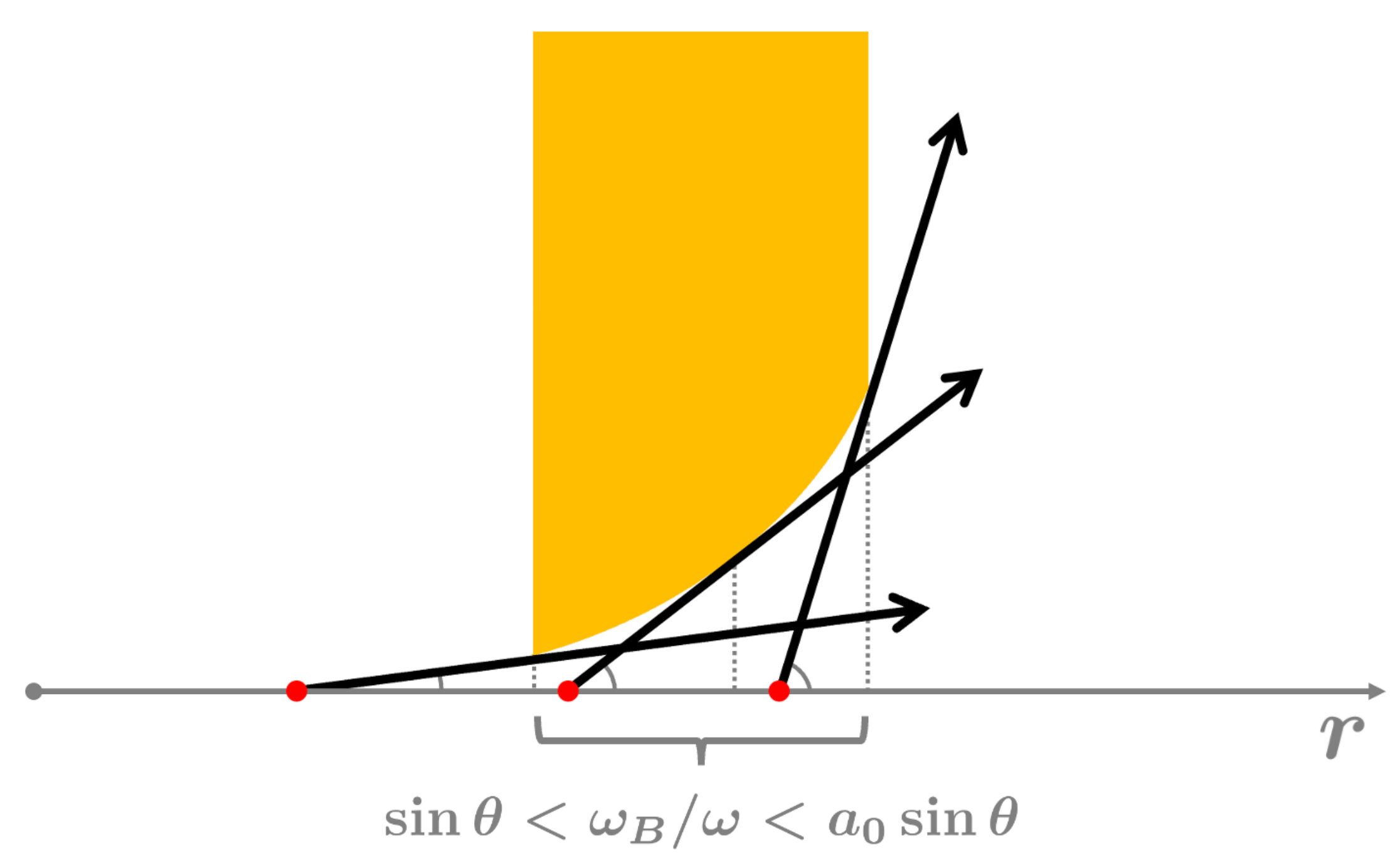}}
		\subfigure{
			\includegraphics[width=0.58\textwidth]{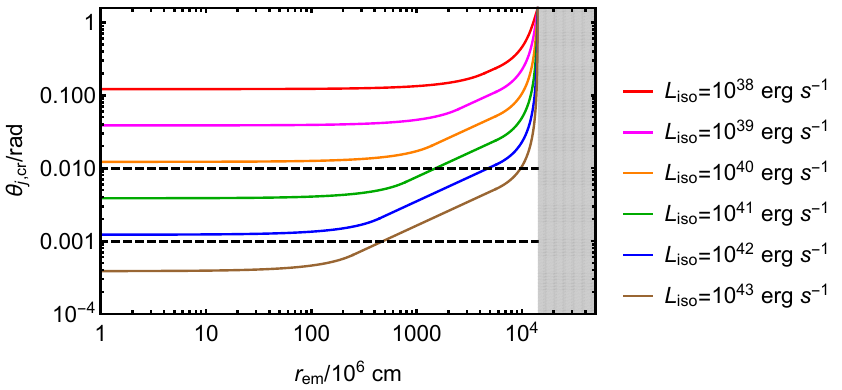}}
		\caption{Three different regimes of constraints on the radiation cone half-opening angle for various emission radii of an FRB in a magnetar magnetosphere (left panel). The orange part represents the region outside the critical transverse spatial size determined by Equation (\ref{lperpmax}). The red dots represent various emission radius of FRBs. The critical radiation cone half-opening angle as a function of the luminosity and emission radius of FRBs (right panel). The two horizontal black dashed lines represent $\theta_{j,\text{cr}}=10^{-3}$ and $10^{-2} \text{ rad}$, respectively. The gray part represents the region outside the light cylinder. The parameters $\xi=10^3$, $\zeta=10^{-3}$, $\nu=10^9\text{ Hz}$, $B_{\text{s}}=10^{15} \text{ G}$, $P=2.95\text{ s}$, and $\theta=10^{-1}\text{ rad}$ are adopted.}
		\label{thetaj}
	\end{center}
\end{figure*}

One intriguing thing is that the finite transverse spatial size can be used to constrain the beaming angle of an FRB. The relationship between the critical radiation cone half-opening angle, the luminosity, and the emission radius of an FRB reads
\begin{equation}
	\begin{aligned}
		\theta_{j,\text{cr}}\left(r_{\text{em}}\right)=\text{min}_{r} \arctan \frac{\mathcal{D}_{\text{cr}}\left(r\right)/2}{r-r_{\text{em}}},
	\end{aligned}
\end{equation}
where $\text{min}_{r}f(r)$ represents the minimum value of $f(r)$ as a function of $r$, and the transverse size of the emitter is neglected. The above expression is valid only when $\sin{\theta}<\omega_B/\omega<a_0\sin{\theta}$ and $r<R_{\text{LC}}$ satisfy. One can obtain three different regimes of constraints on the radiation cone half-opening angle for various emission radii of an FRB in a magnetar magnetosphere, as illustrated in the left panel of Figure \ref{thetaj}.  The critical radiation cone half-opening angle as a function of the luminosity and emission radius of an FRB is shown in the right panel, where we have fixed $\theta=0.1\text{ rad}$ for convenience. For an FRB with isotropic peak luminosity $L_{\text{iso}}\sim 10^{42}\text{ erg s}^{-1}$ and an emission radius $r_{\text{em}}\lesssim 10^{9}\text{ cm}$, the critical half-opening angle of radiation cone $\theta_{j,\text{cr}}\sim10^{-3}-10^{-2}$. A brighter FRB has a more stringent constraint on its beaming angle.

\section{Discussion}\label{discussionsec}

Scattering of FRBs in magnetar magnetospheres was regarded as a criticism toward the magnetospherical origin of FRBs \citep{Beloborodov2021,Beloborodov2022}. It was then pointed out that this scattering effect can be reduced when the propagation direction of an FRB pulse and the background magnetic field line is small or the plasma that an FRB need to penetrate possesses an extremely relativistic outflow \citep{Qu2022,Lyutikov2024}. Later on, \cite{Beloborodov2023} argued again that these conditions cannot be realized in magnetar magnetospheres, even in the open field line region. In this paper, we discarded these presuppositions, and showed that FRBs can still escape from magnetar magnetospheres. We first demonstrated that a small angle between the propagation direction of an FRB and the background magnetic field $\theta\lesssim10^{-2}\text{ rad}$ can naturally mitigate scattering. When the angle $\theta$ is higher, e.g, $\theta\sim 10^{-1}\text{ rad}$, the small transverse spatial size of the FRB pulse can be an important approach to reduce the effect of scattering by particles, and help the FRB escape from the magnetosphere of a magnetar.

If such a mechanism does work, it would impose some potential requirements on the radiation mechanism of FRBs. A straightforward consequence is that the lateral spatial scale of the FRB emitter must be at least smaller than the critical length given by Equation (\ref{lperpmax}), because as an FRB propagates outward, the region it sweeps through will expand. This indicates that the FRB is emitted within a compact region with extremely high energy density in the magnetosphere. Based on this premise, the beaming angle or the emission radius of FRBs can be roughly constrained. A critical radiation cone half-opening angle $\theta_{j,\text{cr}}\sim10^{-3}-10^{-2}\text{ rad}$ is found for an FRB with isotropic equivalent luminosity $L_{\text{iso}}=10^{42}\text{ erg s}^{-1}$ and emitted at a radius $r_{\text{em}}\lesssim10^{9}\text{ cm}$, as shown in the right panel of Figure \ref{thetaj}.

On the one hand, the observations have not strictly constrained the beaming angle of FRBs. From the perspective of magnetospherical models, electrons responsible for generating FRBs typically have a Lorentz factor $\gamma\approx10^2$--$10^3$, which predicts a beaming half-opening angle $\theta_{j}\approx10^{-3}$--$10^{-2}$ \citep{Zhang2020a}. We have marked out these angles in Figure \ref{thetaj}. For the only FRBs associated magnetar SGR 1935+2154, the constraint on the beaming angle for low luminosity bursts ($L_{\text{iso}}\lesssim10^{38}\text{ erg s}^{-1}$) in this paper is very loose ($\theta_{j,\text{cr}}>10^{-2}$). Based on some assumptions, \cite{Chen2023} constrained the half-opening angle $\theta_{j}\lesssim 10^{-2}$ for these bursts, which are consistent with our prediction. On the other hand, the emission radius of FRBs has not been well constrained as well. Our calculation may imply that FRBs could originate at a high altitude, where the ambient plasma density is relatively lower and the nonlinear effect of plasma is stronger. Some models have suggested that FRBs could take place near the light cylinder \citep{Lyubarsky2020,Mahlmann2022,Zhang2022}. A higher emission radius closer to the light cylinder is beneficial to increase the critical transverse size, and therefore naturally alleviate the effect of scattering. It should also be noted that these arguments toward the magnetospheric origin are based on the magnetar parameters empirically inferred from known magnetars \citep{Kaspi2017}. In fact, for some highly active FRB repeaters, it is possible that their sources are a type of magnetar that has not yet been observed. Therefore, our model can be further tested by future observations.

\section{Acknowledgments}

We thank the referee very much for helpful comments that have allowed us to improve our manuscript significantly. We also thank Ze-Nan Liu, Sen-Lin Pang, Yue Wu, and Biao Zhang for their useful discussions. This work is supported by  the National SKA Program of China (grant No. 2020SKA0120302) and National Natural Science Foundation of China (grant No. 12393812).

\normalem
\bibliography{FRB}

\end{document}